\documentclass[conference]{IEEEconf}
\input epsf

\usepackage{graphicx}

\begin{document}

\title{\textbf{\Large GitHub Considered Harmful? Analyzing Open-Source Projects for the Automatic Generation of Cryptographic API Call Sequences\\}}

\author{Catherine Tony$^{*}$, Nicol\'{a}s E. D\'{i}az Ferreyra, and Riccardo Scandariato\\
	\normalsize Institute of Software Security, Hamburg University of Technology, Hamburg, Germany\\
	\normalsize catherine.tony@tuhh.de, nicolas.diaz-ferreyra@tuhh.de, riccardo.scandariato@tuhh.de\\
	\normalsize *corresponding author
}

\maketitle
\begin{abstract}
GitHub is a popular data repository for code examples. It is being continuously used to train several AI-based tools to automatically generate code. However, the effectiveness of such tools in correctly demonstrating the usage of cryptographic APIs has not been thoroughly assessed. In this paper, we investigate the extent and severity of misuses, specifically caused by incorrect cryptographic API call sequences in GitHub. We also analyze the suitability of GitHub data to train a learning-based model to generate correct cryptographic API call sequences. For this, we manually extracted and analyzed the call sequences from GitHub. Using this data, we augmented an existing learning-based model called DeepAPI to create two security-specific models that generate cryptographic API call sequences for a given natural language (NL) description. Our results indicate that it is imperative to not neglect the misuses in API call sequences while using data sources like GitHub, to train models that generate code.
\end{abstract}

\IEEEoverridecommandlockouts

\IEEEpeerreviewmaketitle

\section{Introduction}
\label{sec:intro}
Cryptographic libraries and Application Programming Interfaces (APIs) have become a central component of software-intensive systems as they aim to protect sensitive data against unauthorized access. Despite their importance, studies show that developers struggle with such APIs as many lack the knowledge necessary for their correct usage \cite{cite1, cite3}. This situation often leads to cryptographic APIs being misused by developers, thus exposing software projects to security vulnerabilities. Discussion platforms like Stack Overflow and Stack Exchange are popular among those seeking advice about the proper use of security APIs. Still, the posts available on these platforms frequently showcase wrong examples and insecure code snippets, which makes them faulty sources of security knowledge~\cite{cite14}.

\subsection*{Motivation}

Parametric misuses, in which insecure configurations (e.g., an outdated or weak encryption algorithm) are passed to API method calls, are among developers' most frequent mistakes~\cite{cite9}. Apart from parametric misuses, there are misuses that are induced by incorrect API call sequences that can introduce severe security threats when different API methods are not invoked correctly in the right order. For instance, let us consider an example sequence of API calls made to the Java Cryptography Architecture (JCA) API. JCA is one of Java's cryptographic libraries that provides various cryptographic features. The below example sequence demonstrates the generation of a password-based encryption key using JCA:

\begin{verbatim}
SecureRandom.getInstance 
SecureRandom.nextBytes PBEKeySpec.new 
SecretKeyFactory.getInstance 
SecretKeyFactory.generateSecret 
SecretKey.getEncoded SecretKeySpec.new 
PBEKeySpec.clearPassword
\end{verbatim}

As shown in the example, methods of multiple classes in JCA need to be invoked in a certain sequence to achieve successful password-based encryption. Such a sequence can lead to severe vulnerabilities if the \texttt{clearpassword()} method of \texttt{PBEKeySpec} class is not invoked or if the methods of \texttt{SecureRandom} class (i.e., for the creation of random salt for key specification) are skipped. Despite its importance, information about the adequate implementation approach is often absent or hard to find within the documentation of cryptographic APIs \cite{cite1,cite6}. Moreover, it appears that misuses in method call sequences are relatively less investigated compared to parametric misuses in API calls. Tools like GitHub Copilot\footnote{https://github.com/features/copilot} and Tabnine\footnote{https://www.tabnine.com/} have been recently proposed as means to support developers in automatic code generation. Many of these tools leverage deep learning methods together with data from GitHub's open-source projects to generate automatic code in different programming languages. One could take advantage of these tools to automatically generate secure code that uses different cryptographic APIs. However, to the extent of our knowledge, it is still not clear (i) whether these automatic approaches are trained on secure code, and (ii) if they can generate reliable code for different security use cases such as encryption, digital signature, secret key generation, hashing and so on.

\subsection*{Contribution and Research Questions}

This work seeks to shed some light on the automatic generation of cryptographic API call sequences. Particularly, it analyses the suitability of GitHub data for the implementation of reliable learning models for security code completion. For this, we conducted a two-step analysis of open-source Java projects extracted from GitHub. First, we created a dataset containing cryptographic API call sequences from GitHub and assessed their correctness through existing benchmarks and usage rules. Next, we verified whether the distribution of correct method calls in the dataset is sufficient to train a learning-based model that generates secure cryptographic API call sequences. This second step was conducted using DeepAPI, a deep-learning-based tool developed by Gu et al.~\cite{cite2} that predicts API call sequences for a given natural language annotation or query. We chose DeepAPI because of its demonstrated performance in generating API sequences in Java programming language which is a popularly used language by developers. Since DeepAPI is not trained specifically on cryptographic API usages, we adapted it following transfer learning \cite{transfer-learning} and a dedicated training approach using the new security-specific data which we collected from GitHub in our first step. All in all, we elaborate on the following Research Questions (RQs):

\textbf{\textit{RQ1: }} What are the common misuses present in cryptographic API call sequences from GitHub open-source projects? 

To answer this question, we collected 213 API call sequences from GitHub open-source projects and examined them manually to identify and label the misuses in them. 

\textbf{\textit{RQ2: }} Can a model trained on a large dataset of Java API sequences from GitHub generalize well enough to cover security use cases? 

For this, we took the help of DeepAPI which is trained on millions of generic Java API sequences which may or may not include cryptographic API sequences. We used it to generate sequences for security-specific cases. We then analyzed the correctness and security of the generated sequences.

\textbf{\textit{RQ3: }} Can we apply transfer learning using additional cryptographic API call sequences data from GitHub to improve the security of the generated sequences?

To understand this, we performed transfer learning using the sequences which we collected from GitHub, in the hope of further fine-tuning DeepAPI to generate better cryptographic API sequences. 

\textbf{\textit{RQ4: }} Can we train a model from scratch using a small set of cryptographic API call sequences from GitHub to generate correct call sequences?

For this, we used the base RNN ENcoder-Decoder model for API learning used by DeepAPI and trained it from scratch using only the API sequences collected from GitHub as a part of \textit{\textbf{RQ1}}. We then evaluated the correctness and security of the generated sequences and compared them with the previous results.

Through this study, we were able to have a preliminary understanding of the common misuses in cryptography API call sequences present in GitHub. We have classified these misuses into 6 categories based on their notable characteristics. We also observed that it is imperative to filter out the codes from GitHub, that contain various API sequence misuses, in addition to parametric misuses, before using them for training code generation models.
Detailed descriptions of our approach and findings are presented in the following sections. In the remaining of this paper, the words \textit{cryptographic} and \textit{crypto} will be used synonymously.

\section{Related Work}
\label{sec:rw}
The relevance of cryptography APIs in secure programming and their associated misuses have motivated a significant amount of research. Egele et al.~\cite{cite3} developed a tool called Cryptolint to spot API inconsistencies in Android applications. After inspecting 11,748 apps they found that 88\% of them contained at least one API misuse. Likewise, a study by Hazhirprasand et al.~\cite{cite16} investigated the impact of developers' programming experience on the correct use of cryptography APIs for Java. They observed that 72\% of the analyzed GitHub projects had at least one API misuse. However, these did not correlate with developers' programming experience. 

Wickert et al. \cite{cite9} curated a dataset of parametric API misuses which are extracted from GitHub open source projects. These misuses were identified with the help of a static code analysis tool called CogniCrypt\textsubscript {SAST} \cite{cite5} and through manual inspection. Their dataset consists of 201 parametric misuses. They also classified these misuses into 5 categories based on their underlying cryptographic concepts. They are \textit{transformation, randomization, key, iteration count} and \textit{initialization vector}. Likewise, Hazhirpasand et al.~\cite{cite6} ran the same static code analysis tool, over a collection of Java projects and found that around 75\% of the API calls had security issues. Such misuses were classified thereafter into 5 types: \textit{wrong type}, \textit{wrong object}, \textit{wrong constraint}, \textit{incomplete operation}, and \textit{incomplete order}. As we discuss later in Section~\ref{sec:results}, 2 of these categories (incomplete operation and incomplete order) relate closely to the misuses we have identified in our dataset.

Prior work has also focused on the automatic generation of secure code regarding cryptographic API usage. Such is the case of CogniCryptGen, a tool developed by Krüger et al. \cite{krueger2020}, that generates secure Java code (i.e., calls to cryptographic APIs) for different security use cases. The code generated by the tool follows a set of rules and best practices defined by security experts. Still, and despite its high reliability, CogniCryptGen only covers a reduced number of use cases. A similar approach called FireBugs \cite{firebug2021} detects and repairs cryptography misuses based on security patterns extracted from 5 common parametric mistakes identified in the literature. They are, \textit{weak encryption algorithm, weak hash function, weak encryption algorithm parameter, weak password-based encryption} and \textit{weak random number generator}. 

As mentioned in \ref{sec:intro}, today there are commercial tools available for automatic code generation. GitHub Copilot and Tabnine are two notable ones among them. Copilot is built on top of Codex \cite{codex2021}, which is a large language model trained on millions of publicly available GitHub codes. Similarly, Tabnine is also trained on millions of code from GitHub\footnote{https://www.tabnine.com/}. These tools could be used to generate code that makes use of cryptography APIs to perform different security tasks such as encryption, authentication, and so on. There is an existing paper that discusses the vulnerabilities present in the code generated by Copilot \cite{copilot}. However, to the best of our knowledge, tools like these have not been thoroughly examined under the lens of security when it comes to cryptography API usage. 

\section{Background}

\label{sec:background}

\subsection{DeepAPI}

DeepAPI is a deep learning-based approach that leverages an attention-based RNN Encoder-Decoder mode  to learn API sequences \cite{cite4}. The dataset on which DeepAPI is trained, consists of approximately 7.5 million annotated API sequences in Java expressed as $\langle$API sequence, annotation$\rangle$ pairs (e.g., $\langle$\texttt{Random.new Random.nextInt}, ``generate random number''$\rangle$). Such API sequences are extracted from thousands of Java projects hosted on GitHub using a collection of syntactic rules (c.f.,  \cite{cite2}). Likewise, the corresponding natural-language annotations are obtained from the first line of the corresponding method-level JavaDoc comments. With this training set DeepAPI achieves a BLEU score of 54.42\%\footnote{The BLEU score \cite{bleuscore} is a widely used metric for calculating the similarity between a candidate sequence and a reference sequence. The higher the score is, the closer the candidate sequence is to the reference sequence.}. DeepAPI is one among the very few of its kind with a substantial impact and performance. Hence we have taken advantage of DeepAPI to study the usefulness of crypto API call sequences in GitHub for training learning-based models to generate correct sequences.

\subsection{CrySL}

CrySL \cite{cite5} is a domain-specific language used to specify rules for the correct usage of different APIs. Correct usage may include, i) the order in which the methods of a given API should be called, ii) constraints on parameters passed to API methods, and iii) objects or method calls required to instantiate certain class objects, among others. CrySL considers any deviation from these rules as a misuse of the API. 

\begin{figure}[hbt!]
    \centering
    \includegraphics[scale=.4]{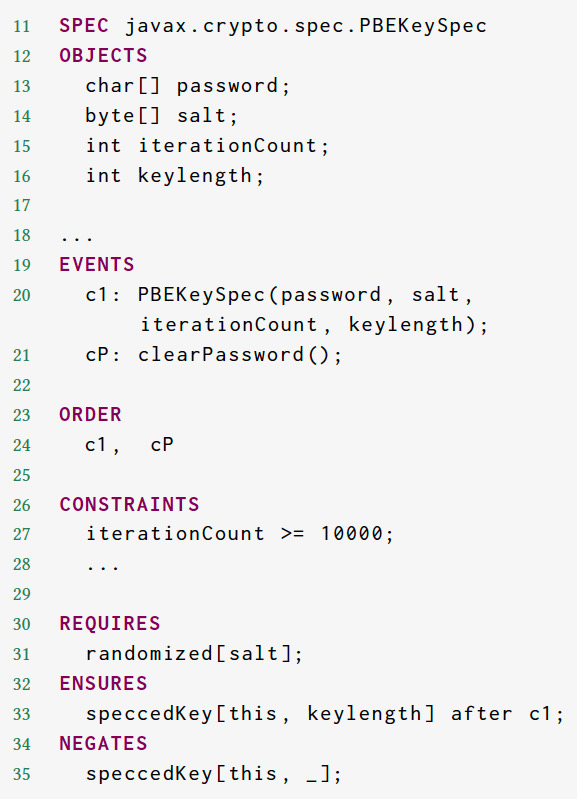}
    \caption{CrySL rule for \texttt{javax.crypto.spec.PBEKeySpec}. Figure borrowed from \cite{cite5}} 
    \label{fig:crysl}
\end{figure}

Figure~\ref{fig:crysl} illustrates a CrySL rule for the JCA class \texttt{javax.crypto.spec.PBEKeySpec}. As it can be observed, the name of the JCA class is specified at the top of the rule followed by an \texttt{OBJECTS} section, which lists the objects and parameters used in it. The \texttt{EVENTS} section specifies the method calls allowed in the class, whereas the order in which such methods should be invoked is defined under the \texttt{ORDER} section. The \texttt{CONSTRAINTS} section enumerates the different constraints on the parameters that are passed to any of the said methods. Prior requirements that need to be satisfied before instantiating a class object (or invoking a class method) are given in the \texttt{REQUIRES} section. The \texttt{ENSURES} section lists the guarantees ensured by the correct implementation of the CrySL rule. Finally, if the rule negates or invalidates any specific property, then such properties are mentioned in the \texttt{NEGATES} section.

Because of its high expressiveness and reliability (rules written by crypto experts), we have used CrySL as one of the main sources to generate the ground truth dataset for this study.

\section{Research Methodology}

\label{sec:methodology}

As mentioned in Section~\ref{sec:intro}, we aim to assess whether GitHub repositories are suitable for training a learning-based model that generates correct crypto API sequences from natural language specifications. For this, we have used DeepAPI, a solution that is not predominantly trained on security tasks but on generic Java code. Furthermore, to improve the correctness of the generated API sequences, we created 2 additional variations of DeepAPI: DeepAPI-plusSec and DeepAPI-onlySec. The former is an extended version of DeepAPI that has undergone transfer learning using our dataset in combination with the original dataset (i.e., the one on which DeepAPI is trained). The latter on the other hand is trained only using our dataset. The following subsections describe our approach in detail.

\subsection{Dataset Creation and Analysis of Misuses in GitHub}
\label{subsec:4A}
\begin{figure}[t]
    \centering
    \includegraphics[width = \linewidth]{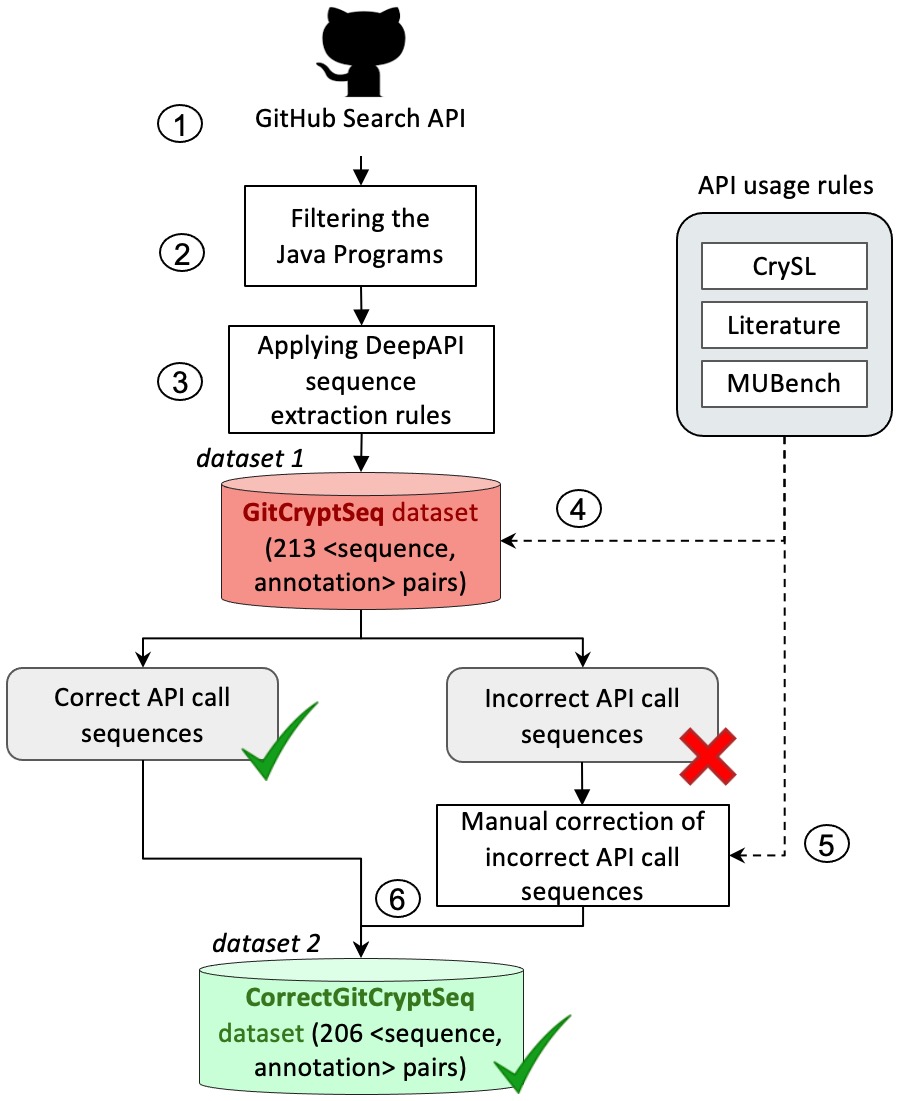}
    \caption{Creation of GitCryptSeq and CorrectGitCryptSeq datasets}
    \label{fig:data-creation}
\end{figure}

We created 2 datasets similar to the one used for training DeepAPI, but instead of having API sequences from the entire JDK library, we focused on cryptography APIs for Java. We concentrated mainly on APIs from JCA and Bouncy Castle\footnote{https://www.bouncycastle.org/} as they are the most popular libraries for cryptography in Java \cite{cite1}. Figure~\ref{fig:data-creation} illustrates the steps we followed for creating both datasets.

\begin{figure}[t]
    \centering
    \includegraphics[width = \linewidth]{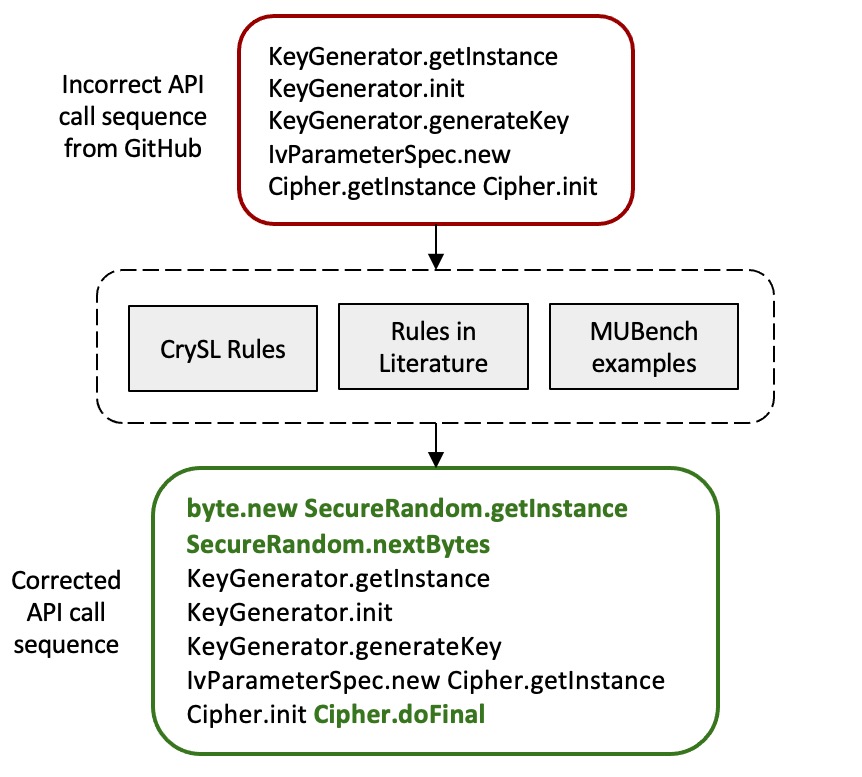}
    \caption{Generation of correct API usage sequence from incorrect sequence}
    \label{fig:workflow}
\end{figure}

\begin{enumerate}
\item Using GitHub's search API, we retrieved a set of Java projects that included calls to the cryptographic APIs mentioned above. We searched specifically for keywords such as \texttt{javax.crypto}, \texttt{java.security}, and \texttt{org.bouncycastle} in their corresponding source code.

\item We retained those projects that (i) implemented a security task using the said crypto libraries, and (ii) included well-readable method-level JavaDoc comments.

\item We extracted the API sequences of each project and their corresponding natural language annotations. We followed the same set of rules used by the creators of DeepAPI \cite{cite2} to retrieve such sequences and their corresponding annotations. Since we performed this step manually, we were also able to create \textit{full API sequences} by following the call chain in multiple methods. These sequences, unlike the ones of DeepAPI, cover complete security tasks defined in JavaDoc comments as DeepAPI restricts its call sequences to individual methods. By the end of this step, 213 crypto API call sequences were included in the first dataset (\textbf{GitCryptSeq}).

\item We used the sequences in GitCryptSeq to generate a second dataset. Particularly, we classified each sequence into \textit{correct} or \textit{incorrect} by applying rules for the adequate use of crypto API. These rules were obtained from the ones defined by CrySL, the current literature \cite{cite3,negResultsCryptoRules2019,cryptoRules2018} along with correct code examples from MUBench \cite{mubench}, an automated benchmark for API misuse detection.

\item Using these same rule sources, we manually converted the incorrect sequences into correct ones. For this, we checked whether there are rules applicable to the API classes included in a particular sequence. If so, we verified if the sequence adheres to them and applied the corresponding changes when necessary. For instance, the sequence shown in Figure~\ref{fig:workflow} consists of 3 crypto API classes: \texttt{KeyGenerator, IvParameterSpec} and \texttt{Cipher}. According to CrySL rules, the  \texttt{IvParameterSpec} and \texttt{Cipher} classes are wrongly used. The former, which initializes vector for crypto API functions, should receive a randomized parameter, whereas the latter requires a call to \texttt{Cipher.doFinal()} for its correct finalization. Hence, the corresponding fix consists of (i) randomizing the parameter passed to \texttt{IvParameterSpec}, and (ii) calling \texttt{Cipher.doFinal()} by the end of the sequence. Sequences with API classes that are not present in any of the three rule sources (i.e., CrySL, literature, and MUBench) were removed from the dataset.

\item Both corrected and already-correct API sequences from GitCryptSeq were combined into a second dataset which we call the \textbf{CorrectGitCryptSeq} dataset.

\end{enumerate}

While correcting the API sequences, we also recorded the number and type of corrections we applied. We observed several misuse patterns and classified them into different categories (\textbf{RQ1}), which are described in Section \ref{subsec:categories}. 

\textbf{Replication Package}: All the data curated for this study are publicly available in a GitLab repository\footnote{Link to data repository: https://github.com/CatherineTony/CryptoAPI-Call-Sequences-Dataset}. 

\subsection{DeepAPI-plusSec and DeepAPI-onlySec}
\label{subsec:4B}
\begin{figure}[hbt!]
    \centering
    \includegraphics[width = \linewidth]{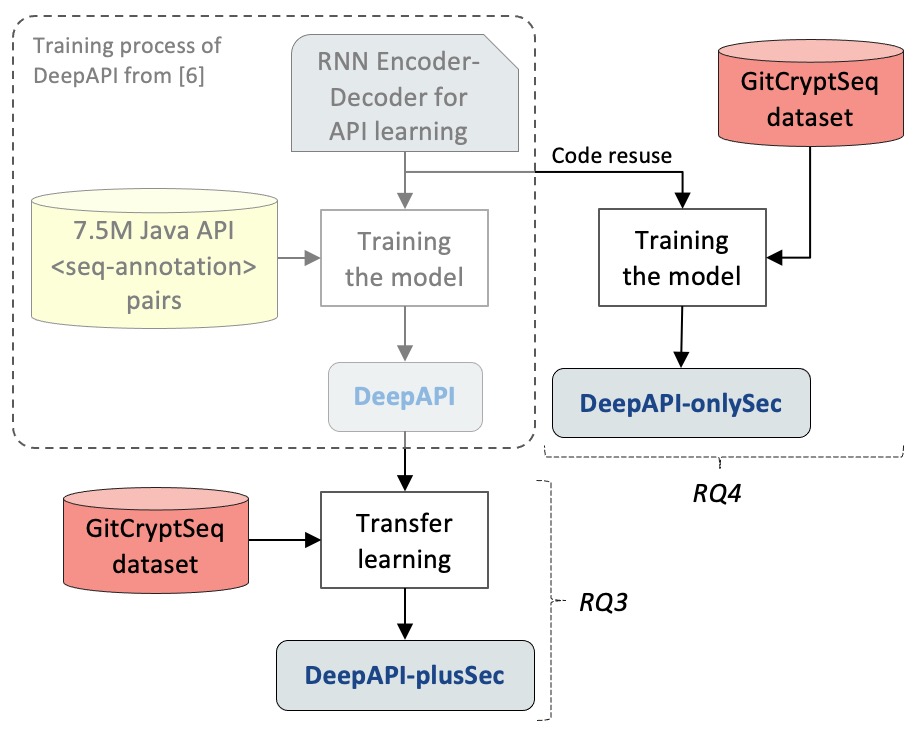}
    \caption{Creating DeepAPI-plusSec and DeepAPI-onlySec from DeepAPI.}
    \label{fig:variations}
\end{figure}

In addition to the DeepAPI evaluation (\textbf{RQ2}), we assessed the performances of 2 variations of it, targeting the generation of secure/correct crypto API sequences from NL descriptions using the data from GitHub. These 2 variations are called DeepAPI-plusSec and DeepAPI-onlySec, and their differences are depicted in Figure~\ref{fig:variations}. 

The part of the diagram that is covered by the dotted square shows how DeepAPI is originally trained according to \cite{cite2}. We created DeepAPI-plusSec to understand the effect of transfer learning using the data from GitHub on the performance of DeepAPI for security-related scenarios. For that, we used the sequences and annotations of GitCryptSeq to further adapt the model for security purposes via transfer learning. DeepAPI's neural networks are trained using sequence and annotation vocabularies each containing the top 10,000 most frequent words in the dataset. However, these vocabularies omit many classes and method names linked to JCA and BouncyCastle as their frequency inside DeepAPI's dataset is very low. To rectify this we searched all unique words in the sequences and annotations inside GitCryptSeq and kept those that were not included in the original vocabulary. We used these words to enrich the original vocabulary and used them in combination with GitCryptSeq for re-training DeepAPI into \textbf{DeepAPI-plusSec}.

To further understand whether the GitHub-retrieved sequences are suitable for training a security-aware deep learning model (\textbf{\textit{RQ4}}), we created two new vocabularies: one containing API sequences and another one with their corresponding NL annotations. Both vocabularies consist of 219 unique words extracted from the GitCryptSeq dataset. We used these vocabularies to build \textbf{DeepAPI-onlySec}, a new version of DeepAPI targeting the automatic generation of secure crypto API sequences from NL queries. For this, we trained DeepAPI's RNN Encoder-Decoder model from scratch and only with the 213 sequence-annotation pairs inside GitCryptSeq. 

According to \cite{cite2}, the best performance of DeepAPI is achieved using RNNs with 1000 hidden units per layer and setting the dimension of word embeddings to 120. We followed this same approach when setting the hyperparameters of DeepAPI-plusSec and DeepAPI-onlySec without further tuning.

\subsection{Evaluating the Crypto API Sequence Generation Models}

\begin{figure}[hbt!]
    \centering
    \includegraphics[width = \linewidth]{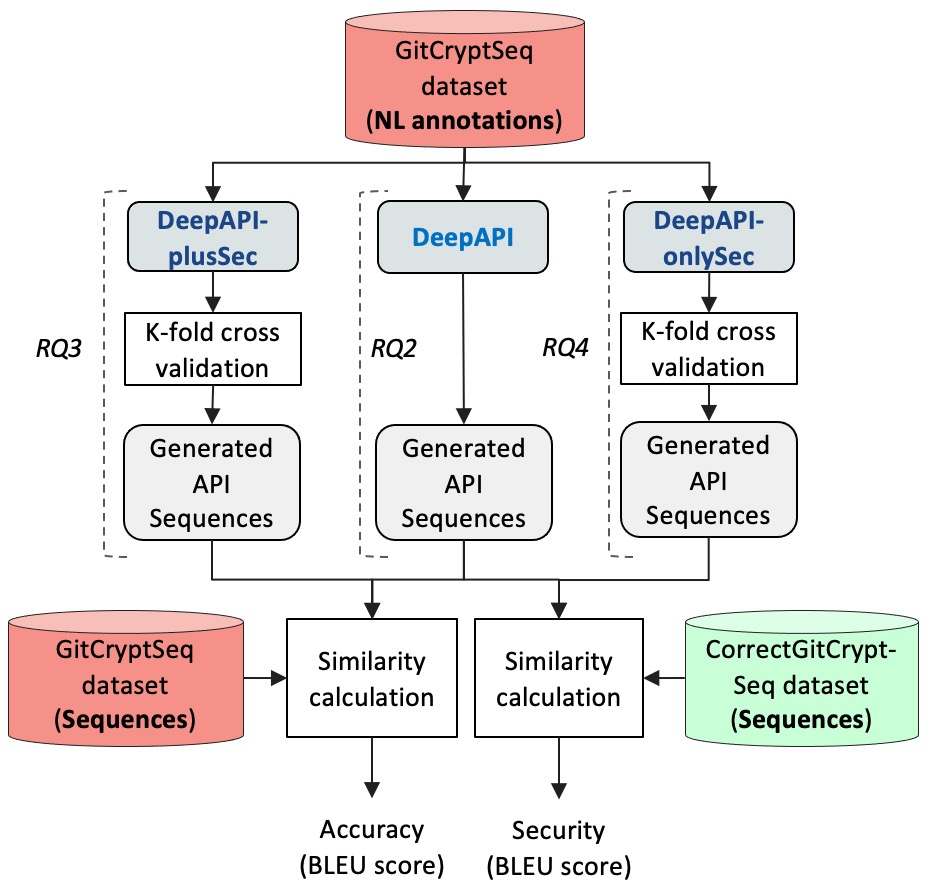}
    \caption{Process followed to evaluate the performance of DeepAPI, DeepAPI-plusSec, and DeepAPI-onlySec in terms of accuracy and security}
    \label{fig:pred-process}
\end{figure}

We used GitCryptSeq and CorrectGitCryptSeq to evaluate the performances of DeepAPI, DeepAPI-plusSec, and DeepAPI-onlySec. This process is depicted in Figure \ref{fig:pred-process}.

We used all the NL queries in GitCryptSeq as the test set for assessing DeepAPI (\textbf{\textit{RQ2}}). Since DeepAPI-plusSec (\textbf{\textit{RQ3}}) and DeepAPI-onlySec (\textbf{\textit{RQ4}}) were trained on a relatively small dataset, we used k-fold cross-validation to evaluate their performance (k=10). We combined the API call sequences generated for each fold and evaluated them together. 

We performed 2 types of performance evaluation. First, we measured each model's accuracy by calculating the similarity between the generated API call sequences and the corresponding sequences in the GitCryptSeq dataset. We chose the BLEU metric for computing sequence similarity, as this was also used during DeepAPI's performance evaluation. A detailed explanation of the BLEU score and its application to DeepAPI is available in \cite{cite2}. We have followed the same method for this study. 

The second evaluation involved assessing the security of the generated sequences using our ground truth dataset CorrectGitCryptSeq. This dataset includes the sequences from GitCryptSeq that we manually examined and corrected. We calculated the similarity between the sequences generated by the three models and the corresponding sequences in the CorrectGitCryptSeq dataset using the BLEU metric. The similarity calculation was performed only for correct and secure sequences available in CorrectGitCryptSeq. The results of our evaluation are presented and discussed in the next 2 sections.  

\section{Results}
\label{sec:results}

\subsection{Datasets Summary}
The GitCryptSeq dataset contains 213 crypto API call sequences and the corresponding NL annotations extracted from 42 GitHub open-source projects. These projects were selected from the first 100 result pages obtained through GitHub search API using the keywords listed in Section \ref{sec:methodology}. They all contain Java code that uses either JCA or BouncyCastle and includes readable JavaDoc comments (NL annotations). The projects in our dataset have an average of 5.07 crypto API sequences. We were able to find correct usage rules for 206 out of 213 API sequences. These 206 sequences were corrected and included in the CorrectGitCryptSeq dataset.

\subsection{Categorization of Misuses in API Usage Sequences}
\label{subsec:categories}

To answer (\textbf{\textit{RQ1}}) we analyzed the sequences in the GitCrytpSeq dataset to identify API misuses. We used the BLEU score to determine the similarity between the 206 crypto API sequence extracted from GitHub and the corresponding correct sequence inside the CorrectGitCryptSeq dataset. The average BLEU score was 80.14\% and 83 of the sequences achieved a BLEU score of 100\%. This means that 59.7\% of the sequences under analysis had at least one API misuse. 

\begin{figure}[hbt!]
    \centering
    \includegraphics[width=\linewidth]{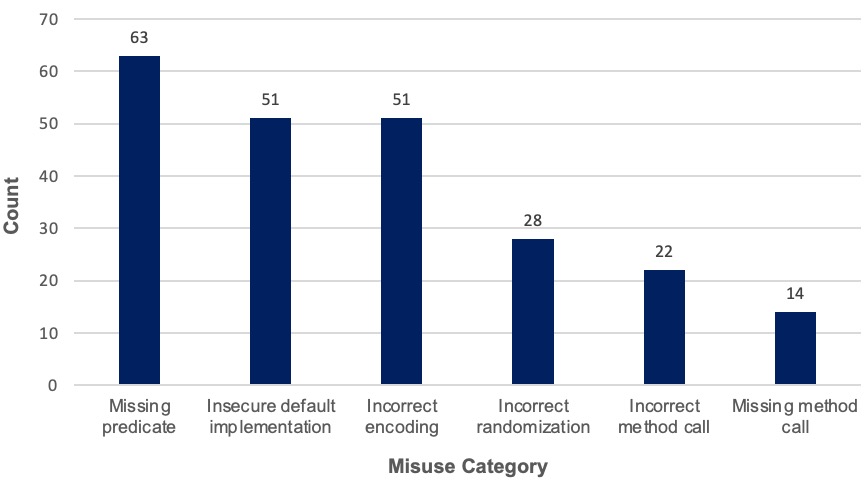}
    \caption{Distribution of the misuse categories in our GitCryptSeq dataset} 
    \label{fig:misuses}
\end{figure}

We observed some common misuses in the analyzed API sequences and classified them into 6 categories: \textit{missing predicate, insecure default implementation, incorrect encoding, incorrect randomization, incorrect method call} and finally \textit{missing method call}. Their occurrence frequencies in our dataset are shown in Figure~\ref{fig:misuses}. 

\subsubsection{Missing Predicate}
This is caused when a necessary requirement for the initialization or usage of a certain API class object is not fulfilled. For our analysis, the necessary requirements were mostly obtained from the \texttt{REQUIRES} section (Figure \ref{fig:crysl}) of the CrySL rule written for the corresponding API class. For instance, \texttt{Cipher} is a JCA class that enables encryption and decryption through a cryptographic cipher. The CrySL rule \cite{cite5} for this class states that, for the secure instantiation of a \texttt{Cipher} object, it requires a generated \texttt{SecretKey, PublicKey} or \texttt{PrivateKey} object using a given algorithm. Such a key object can be obtained through the different method calls of \texttt{SecretKeyFactory}, \texttt{KeyGenerator}, \texttt{KeyPairGenerator} or \texttt{SecretKeySpec} classes depending on the encryption needs. API usage sequences that do not instantiate class objects (or methods) following these requirements are considered as missing predicate misuses.

\subsubsection{Insecure Default Implementation}

\begin{figure}[hbt!]
    \centering
    \includegraphics[width=\linewidth]{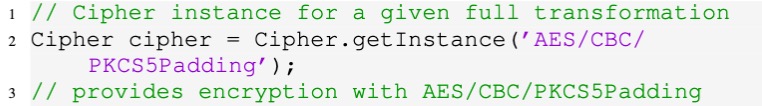}
    \caption{Secure Cipher Instance with full transformation and no provider}
    \label{fig:code1}
\end{figure}

\begin{figure}[hbt!]
    \centering
    \includegraphics[width=\linewidth]{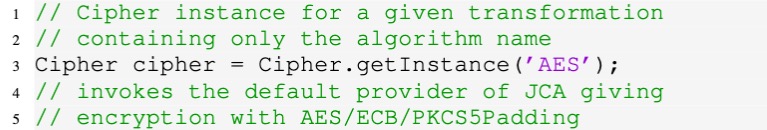}
    \caption{Insecure Cipher Instance with incomplete transformation and no provider}
    \label{fig:code2}
\end{figure}

\begin{figure}[hbt!]
    \centering
    \includegraphics[width=\linewidth]{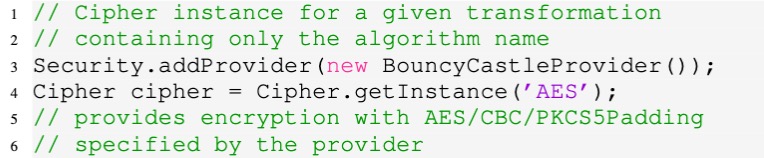}
    \caption{Secure Cipher Instance with incomplete transformation and additional provider}
    \label{fig:code3}
\end{figure}

Some API classes allow users to create objects with default parameters. However, defaults do not necessarily lead to secure instances of crypto API classes. Figures \ref{fig:code1}, \ref{fig:code2} and \ref{fig:code3} show 3 different ways in which the APIs sequences in our dataset create \texttt{Cipher} objects. In principle, a \texttt{Cipher} object can be created by passing a ``transformation'' parameter to the \texttt{getInstance()} method. According to the JCA documentation, such a parameter is an operation to be performed on an input. It comprises the name of the encryption/decryption algorithm, the mode of encryption/decryption, and the padding scheme used. Among these, the algorithm name is a mandatory entry for a transformation while the other two are optional. However, the security of a cipher cannot be guaranteed only through a secure algorithm. Using an appropriate mode and padding scheme is also important.


Figure \ref{fig:code1} shows the creation of a \texttt{Cipher} object with a transformation that includes algorithm name, mode and padding scheme. The Advanced Encryption Standard (AES) algorithm combined with a Cipher Block Chaining (CBC) mode and a PKCS5Padding padding scheme is considered to be a secure implementation for performing encryption/decryption provided a separate message authentication code (MAC) is generated by the sender after encryption and is verified by the receiver before decryption \cite{cite7}. On the other hand, the code in Figure \ref{fig:code2} creates the object by passing a transformation that contains only the algorithm name (i.e., AES). Hence, the \texttt{Cipher} is instantiated with JCA's default parameters. That is, using Electronic Code Book (ECB) and PKCS5Padding as the default mode and padding scheme, respectively. AES is considered a very secure algorithm for cryptography \cite{cite8}. However, AES combined with ECB is considered to be unsafe for the encryption of blocks larger than 128 bits. An alternate way to rectify this is to specify the provider before creating the object as shown in Figure \ref{fig:code3}. Here, ECB is replaced by CBC, a more secure operation mode, by adding a \texttt{BouncyCastleProvider} to the \texttt{Cipher} object. Thus, we considered those objects created without a full transformation and without calling a specific provider as misuses. More broadly, any misuse that is caused by default implementations in JCA and BouncyCastle falls under the \textit{insecure default implementation} category.

\subsubsection{Incorrect Encoding}
Other frequent errors observed in our dataset relate to the conversion of \textit{String} values to bytes arrays. API classes like \texttt{Cipher}, \texttt{MessageDigest}, \texttt{HMac}, etc.  include methods that require parameters in the form of byte arrays. These are often obtained by converting a string data value into a byte array. In our dataset, a common method used to perform this conversion is \texttt{String.getBytes()}. However, if this method is invoked without specifying an encoding format, it could result in hash values and output streams with inconsistent formats, leading to faulty encryption or data integrity checks. One way to correct this issue is to ensure that the encoding format is specified in the method call (e.g., \texttt{String.getBytes(StandardCharsets.UTF\_8)}). Alternatively, one could use \texttt{Strings.toUTF8 ByteArray()} from BouncyCastle to convert a \textit{String} to an array of bytes in UTF8 format. 

\subsubsection{Incorrect Randomization}

As the name indicates, the misuses that come under this category are caused due to improper or missing randomization of values passed to the API instantiation methods. For example, the \texttt{IvParameterSpec} class of JCA which is used to specify an initialization vector for creating ciphers requires a randomized byte array. According to the CrySL rules, the randomization of this byte array can be achieved through proper calls to the methods of \texttt{SecureRandom} class.

\subsubsection{Incorrect Method Call}

To perform a cryptographic task, several method calls have to be made from multiple crypto API classes. When a wrong method is called according to the rules written in \texttt{REQUIRES} and \texttt{ORDER} section of CrySL and other existing literature, then it is counted as a misuse. 
To give an example, in order to perform symmetric encryption, a secret key is required. One way to do this is to generate a key using \texttt{SecretKeyFactory} class, with a given key specification. The specification could be regarding the algorithm used to generate the key or the encoding format of the key and can be generated using JCA classes such as \texttt{DESKeySpec, PBEKeySpec} etc. However, if Data Encryption Standard (DES) algorithm is used to construct or specify this key, it will lead to high-security threats from brute-force attacks \cite{cite9}. Thus, when \texttt{DESKeySpec} is used instead of \texttt{PBEKeySpec} or other secure specifications, it is considered as an incorrect method call. 

\subsubsection{Missing Method Call}
CrySL dictates the order in which the methods of the same API class should be called. It also gives information on which method calls are mandatory and which of them are optional. A misuse is recorded when one of the mandatory methods of a class is not called by an API sequence of our dataset. For example, while using \texttt{PBEKeySpec} for password-based data encryption, developers sometimes forget to call the \texttt{clearPassword()} which can cause security threats \cite{cite5}.

\subsection{Performances of DeepAPI, DeepAPI-plusSec and DeepAPI-onlySec}
\label{results:b}
As mentioned in Section \ref{sec:background}, DeepAPI is trained on millions of annotated API sequences in Java while DeepAPI-plusSec is the result of performing transfer learning on DeepAPI using GitCryptSeq dataset that contains 213 crypto API sequences and annotations. On the other hand, DeepAPI-onlySec is trained only on the GitCryptSeq dataset.

\begin{figure}[hbt!]
    \centering
    \includegraphics[width=\linewidth]{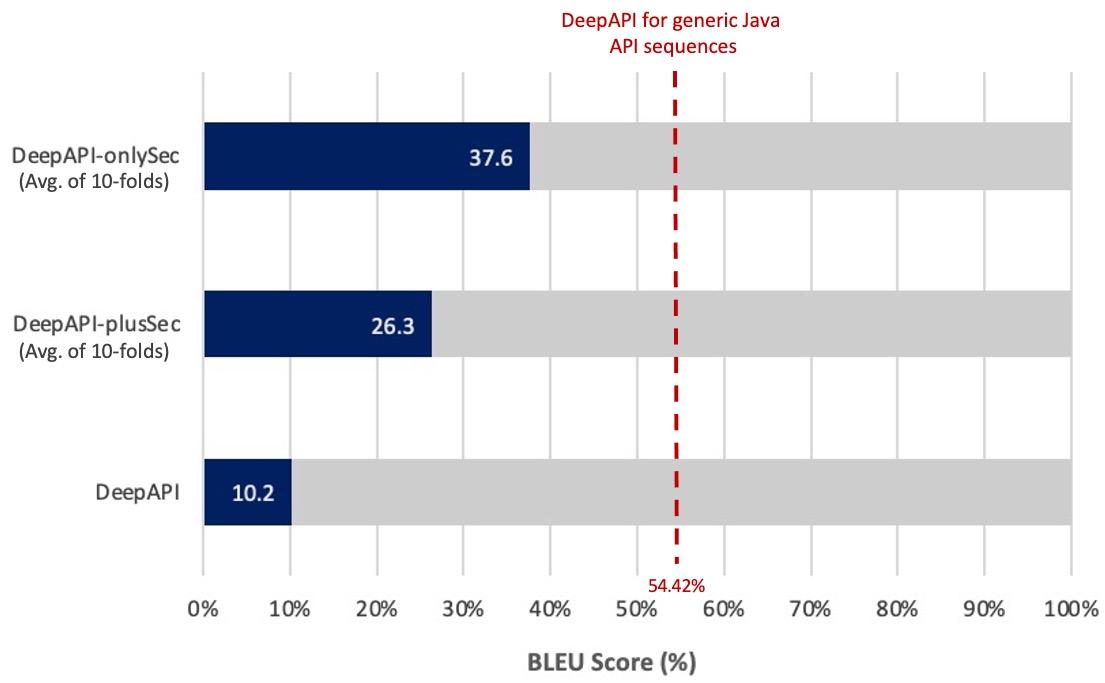}
    \caption{BLEU score (\%) between target sequences in GitCryptSeq dataset and the generated sequences by DeepAPI, DeepAPI-plusSec and DeepAPI-onlySec} 
    \label{fig:predictions}
\end{figure}

The performance accuracy of DeepAPI was measured using the 213 sequences in our dataset as the test set. DeepAPI generated API sequences for all the NL annotations in the GitCryptSeq dataset and we measured their BLEU score to assess the quality of the generated sequences (\textbf{\textit{RQ2}}). On the other hand, the performances of DeepAPI-plusSec and DeepAPI-onlySec (\textbf{\textit{RQ3, RQ4}}) were measured through 10-fold cross-validation because of the limited size of our GitCryptSeq dataset. The accuracy of the sequences generated by DeepAPI, DeepAPI-plusSec, and DeepAPI-onlySec are shown in Figure \ref{fig:predictions}. The red dotted line in the figure indicates the performance of DeepAPI that was reported for generic Java API sequences in \cite{cite2}. None of the models managed to reach the level of performance that DeepAPI achieved for generic Java sequences. However, when tested against crypto API sequences, DeepAPI delivered the least performance (10.2\% BLEU score) out of the three models. DeepAPI-onlySec on the other hand, was able to attain a BLEU score of 37.6\%, becoming the best-performing model among the three. 

\begin{figure}[hbt!]
    \centering
    \includegraphics[width=\linewidth]{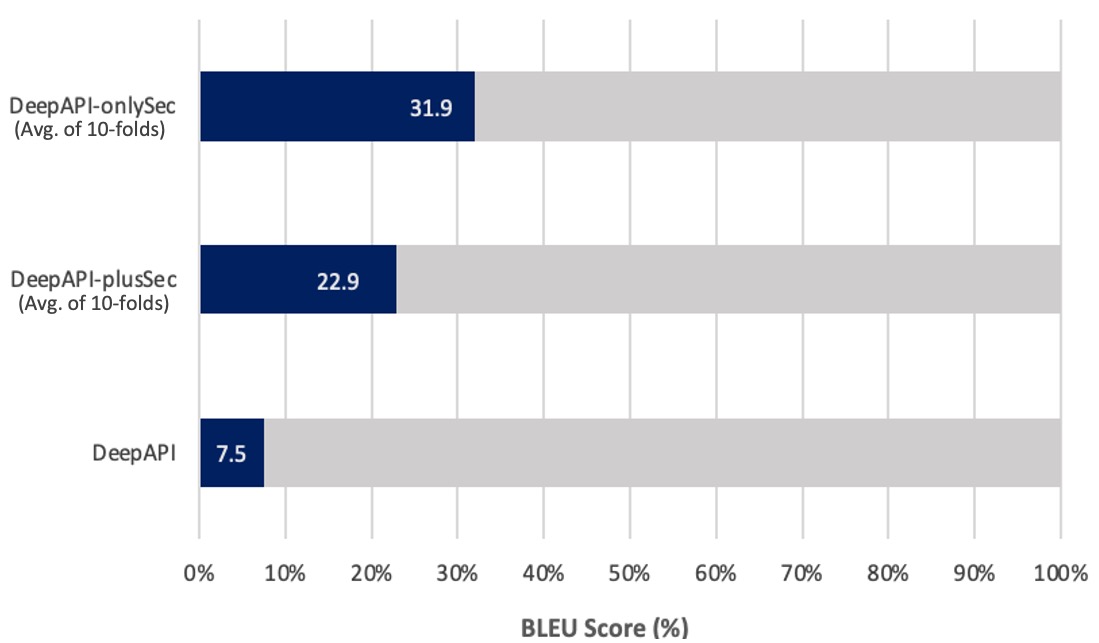}
    \caption{BLEU score (\%) between the correct sequences from CorrectGitCryptSeq dataset and the generated sequences by DeepAPI, DeepAPI-plusSec and DeepAPI-onlySec respectively} 
    \label{fig:correctness}
\end{figure}

Next, we analyzed the security of the API sequences generated by DeepAPI, DeepAPI-plusSec, and DeepAPI-onlySec by measuring their similarity to the corresponding sequences in the CorrectGitCryptSeq dataset (i.e., via the BLEU metric). As mentioned earlier, we considered only those 206 sequences with an entry in the CorrectGitCryptSeq dataset for this measurement. The results are as shown in Figure \ref{fig:correctness}. Similar to the previous case of performances, DeepAPI-onlySec has managed to generate sequences with the highest level of security whereas the sequences generated by DeepAPI are the least secure among the three models.

\section{Discussion}
\label{sec:discussion}

\subsection{Common Misuses in API Usage Sequences from GitHub (\textit{\textbf{RQ1}})}

As mentioned in Section \ref{sec:results}, the crypto API sequences from GitHub attained a BLEU score of 80.14\% against the corresponding safe and correct sequences. According to an evaluation conducted by Lavie et al. \cite{cite11}, a BLEU score above 30\% is considered ``understandable'' and scores over 50\% can be seen as ``very good''. Considering this, a similarity score of 80.14\% seems impressive. However, it does not necessarily mean that the crypto API sequences from GitHub are safe enough to be reused in software projects. One of BLEU's  drawbacks is that, when it comes to code sequences, it may assign small penalties to minor/superficial changes that can alter significantly a code's behavior \cite{cite12}. After computing the number of misuses present in each of the API sequences from GitHub we found out that 59.7\% of the sequences under consideration had at least one misuse.


As shown in Section \ref{subsec:categories}, the most common misuse type is the \textit{missing predicate}. The JCA and BouncyCastle documentations often fail to provide useful insights about the necessary object instantiations and method calls necessary before performing an action using one of their classes \cite{cite6}. Based on our attempts to find the correct order in which the methods and the objects of these APIs should be used, we have observed that there is a shortage of reliable and understandable sources of information. To the best of our knowledge, aside from the JCA documentation itself, CrySL rules are one of the most reliable sources specifying clear requirements and the order in which the objects and methods of these classes should be used. Another alternative is to check the correct usage examples in the MUBench dataset. However, from this dataset one can only infer correct usage requirements by examining multiple examples of correct implementations performing the same task. In other words, the rules to be followed are not explicitly mentioned in this dataset. A third option would be browsing through various research papers to gather these rules \cite{cite3}\cite{negResultsCryptoRules2019}\cite{cryptoRules2018}. Using all the sources mentioned above for generating correct API sequences is time-consuming and requires significant effort. Due to this reason, one may settle with a code snippet performing a cryptography task in a functionally correct way. This could be a reason behind the missing predicate misuse and its high frequency in our dataset.

Insecure default implementation is the second most common misuse type in our dataset, and it was mainly observed in the creation of \texttt{Cipher} objects with incomplete transformations. AES is considered a secure algorithm for cryptography \cite{cite8} and it may be the reason why many projects in our dataset use it instead of DES (this last one is considered insecure). However, many developers seem to ignore that using the standard JCA provider for creating \texttt{Cipher} instances can lead to serious issues. This is because the default encryption mode of this provider is ECB which can make AES also susceptible to various security threats. 

The incorrect encoding misuse in our dataset is as prevalent as the insecure default implementation misuses. Such a misuse leads to false negatives for data integrity checks. Still, compared to the other types, it can be considered less severe as it does not lead to serious security attacks at the moment. On the other hand, risks caused by incorrect randomization misuse are very straightforward. Insufficient randomization can enable attackers to guess parameters with considerably less effort and, in the worse case, allow them to break the encryption. 

Two types of incorrect method calls were observed in our dataset. The first and most common one is the use of \texttt{DESKeySpec}, which can bring serious consequences as DES is considered highly insecure. The second (though less frequently observed) is when methods for generating wrong keys are used for asymmetric encryption (i.e., generating public keys where private keys are needed and vice versa). These mistakes seem accidental and less severe due to their nature and low frequency in the dataset. 

The least common misuse type in our dataset is the missing method call. Particularly, of methods like \texttt{PBEKeySpec.clearPassword} and \texttt{Cipher.doFinal}. The \texttt{clearPassword} method of \texttt{PBEKeySpec} is used to remove the internal copy of the password used for password-based encryption from the system. Without this method, the password will be stored in the system for a prolonged period of time making it accessible for developers or attackers leading to different security threats. Nevertheless, the exclusion of this method call does not affect the functionality of the encryption process. This could be a reason why developers tend to forget or ignore this method call. On the other hand, the \texttt{doFinal} method from \texttt{Cipher} class is necessary to successfully complete the encryption/decryption operation. This may explain the less number of occurrences of this misuse in our dataset.

To summarize, through a closer look at the sequences from GitHub, we conclude that, despite having a high BLEU score with the correct and secure crypto API sequences, more than half of the sequences in our dataset contain at least one of the above-mentioned misuses. 

\subsection{Training Models Using API Sequences from GitHub (\textit{\textbf{RQ2, RQ3, RQ4}})}


DeepAPI, which is pre-trained on millions of generic Java sequence-annotation pairs from GitHub, obtained a BLEU score of 10.2\% when tested against crypto API sequences performing various cryptographic tasks. However, the creators of DeepAPI reported a BLEU score of 54.42\% when tested for generic Java API sequences. To further understand this performance decrease when it comes to security tasks, we checked the number of sequences present in the test set of DeepAPI. That is, those containing at least one API class from JCA or BouncyCastle. We found that, out of 10,000 API sequences in the DeepAPI test set, only around 2\% had a call to any of the crypto API classes under consideration. This indicates a considerable shortage of cryptographic API sequences in the DeepAPI dataset. 

The vocabularies under which DeepAPI is trained consist of the 10,000 most frequent words in the API sequences and their corresponding NL language annotations. On the other hand in the GitCryptSeq dataset, the vocabularies of the API sequences and annotations have 219 unique words each.  
Out of these 219 words, only 125 are present in the API sequences vocabulary of DeepAPI. In the case of NL annotations, only 175 out of 219 words are present in DeepAPI's annotations vocabulary. After analyzing the results file created by DeepAPI 
it was evident that it did not recognize many of the words from the GitCryptSeq dataset. To answer \textbf{\textit{RQ2}}, in spite of having collected 7.5 million API sequences from GitHub, DeepAPI lacked sufficient data associated with security scenarios and consequently the necessary vocabulary. This could have contributed to the low performance of DeepAPI in this case. The similarity between the generated sequences by DeepAPI and the corresponding correct and secure sequences in the CorrectGitCryptSeq dataset was measured using the BLEU score. It has measured to 7.5\% which is very poor at the same time expected considering the model's performance in different security scenarios.


The 26.3\% BLEU score of DeepAPI-plusSec (\textbf{\textit{RQ3}}) is not a good score either \cite{cite11}. However, its prediction accuracy doubled when compared to DeepAPI after transfer learning. We think that such a performance can be further improved by increasing the size of our dataset. DeepAPI-plusSec obtained a BLEU score of 22.9\% when measured against CorrectGitCryptSeq dataset. Furthermore, we can observe that the security of the generated sequences is almost three times higher than the ones obtained with DeepAPI. This is not surprising as the sequences in the GitCryptSeq dataset have a similarity score of 80.14\% with regard to the CorrectGitCryptSeq dataset. Hence, as the prediction accuracy of the model increases, the measured security will also increase.

Although transfer learning is seen as efficient when there is a shortage of data, some limitations must be acknowledged \cite{cite15}. One of the challenges of transfer learning is that, if there is a significant mismatch between the domain of the dataset used in the pre-trained model and that of the fine-tuned model, it can lead to negative transfer or limited performance \cite{cite15}. To investigate if this played a role in the DeepAPI-plusSec's performance, we further analyzed the two datasets and checked whether there are any distinctive differences between the crypto API call sequences and generic Java API call sequences. After calculating the average length (number of method calls) of the sequences in each of the datasets, we found that there is no big difference in length between the sequences used to train DeepAPI (7.11 method calls) and the sequences used to perform transfer learning on DeepAPI-plusSec (8.57 method calls). 

We also computed the number of sequences in each set having more than 7 method calls. We chose 7 because 7.11 is the lowest average length among the 2 datasets. We observed that 61\% of the sequences in the GitCryptSeq dataset had more than 7 method calls whereas only 24.5\% in the original dataset used by DeepAPI contain more than 7 method calls. Even though the average length of the sequences in both datasets is about the same, one can see 
 that there is a substantially more number of longer sequences in the GitCryptSeq dataset compared to the dataset used by DeepAPI.
 Such a dissimilarity in the data distribution could be the reason why the performance improvement of DeepAPI-plusSec after transfer learning was limited. Nevertheless, the performance did increase after transfer learning showing negligible signs of negative transfer as such.

Due to the limitations of transfer learning and to answer \textbf{\textit{RQ4}}, we trained the RNN Encoder-Decoder model of DeepAPI from scratch using the data in GitCryptSeq dataset (DeepAPI-onlySec). From the results in section \ref{results:b}, we see that DeepAPI-onlySec was able to achieve a BLEU score of 37.6\%, improving the one of DeepAPI-plusSec (26.3\%). Nevertheless, we believe that 213 data samples are still not enough to properly tune a model that is pre-trained on 7.5 million data points. This could be the reason for DeepAPI-onlySec's relatively increased performance, as it has not seen any data other than the cryptography API sequences in our dataset. As for the security of the generated sequences, DeepAPI-onlySec (BLEU score: 31.9\%) performs almost 5- and 1.6-times better than DeepAPI and DeepAPI-plusSec (BLEU score: 19.2\%), respectively. Answering \textbf{\textit{RQ4}}, we observe that the correctness and security of the generated sequences do benefit from dedicated or domain-specific training.  We did not analyze the security of the generated sequences by the three models beyond the BLEU score because the maximum similarity score obtained by the models did not exceed 31.9\%. All in all, assessing the type and number of misuses present in the generated sequences by these models with such a limited similarity score is inconsequential. 

\section{Threats to Validity}
Although this study has yielded some interesting results, there are some limitations that need to be acknowledged. 
While creating our dataset, the popularity (number of stars, forks, or downloads) of the open-source projects was not considered as a selection criterion since our focus is to identify the misuses present in GitHub as a whole and not in a selected subset of high-quality projects. The motivation behind this reasoning comes from the fact that some large code generation models such as Codex \cite{codex2021}, do not filter out projects with low popularity in terms of the number of stars, forks, or downloads.

The datasets containing crypto API usages used in this study contain a relatively small number of data samples. Hence the categorization of common misuses in crypto API call sequences found in GitHub open-source projects should be seen as a preliminary attempt. Furthermore, DeepAPI-plusSec and DeepAPI-onlySec are trained using our small datasets which may have led to biased and limited performances. Nevertheless, we have performed k-fold cross-validation to rectify these effects to an extent. Hence, we consider the results and inferences obtained in this study to be meaningful and motivate us to expand the dataset to improve the learning-based models.

Finally, the identification of misuses and the generation of correct call sequences were done manually which may challenge the correctness of the datasets. However, we have adhered strictly to the well-defined usage rules written in CrySL, existing literature, and the sequences in MUBench in performing these tasks. Hence, the degree of error in our work has been largely minimized.

\section{Conclusion}

\label{sec:conclusion}

One of the main goals of this study was to have a preliminary understanding of misuses associated with crypto API call sequences in GitHub open-source projects. By examining a sample of such call sequences from GitHub, we identified 6 types of most commonly occurring sequential crypto API misuses. We also delved deep into the severity and extent of these 6 misuse types and explored the reasons leading to their occurrence in open-source projects.  

Additionally, we also evaluated the performances of DeepAPI model as well as 2 variations of this model (DeepAPI-plusSec and DeepAPI-onlySec) for generating crypto API call sequences. From the results of DeepAPI, DeepAPI-plusSec, and DeepAPI-onlySec, we understand that training a learning-based model using general programming data from GitHub is not sufficient to cover security-related scenarios using different cryptography APIs. More domain-specific training using correct data is required to address the security aspects of programming. Today, there are various learning-based models that can automatically generate code such as GPT-3 \cite{gpt-3}, Codex, PaLM \cite{palm}. They achieve this by training on a vast amount of programming and natural language data available online. In addition to evaluating the functional correctness of the code generated by such models, it is essential to verify that the generated code correctly implements different security concepts that use complex cryptographic libraries.

For future work, we intend to create a bigger dataset consisting of correct crypto API call sequences and standardize it in order to enable better security domain-specific training of learning models. Currently, the models in this paper can only generate call sequences in Java programming language. Hence, it would be interesting to train these models with sequences from other prominent programming languages and evaluate the results. 
We also plan to compare the prevalence and severity of parametric and sequential misuses of crypto APIs in code generated by large language models for different programming languages that leverage GitHub open-source projects for generating code.


\section*{Acknowledgment}
This work was partly funded by the European Union's Horizon 2020 programme under grant agreement No. 952647 (AssureMOSS).




%

\end{document}